\documentclass[prl,aps,twocolumn,showpacs]{revtex4}

\usepackage{pxfonts}
\usepackage{graphicx}
\begin{document}

\title{Pancharatnam-Berry phase in condensate of indirect excitons}

\author{J.\,R.~Leonard, A.\,A.~High, A.\,T.~Hammack, M.\,M.~Fogler, L.\,V. Butov}
\affiliation{Department of Physics, University of California at San Diego, La Jolla, California 92093-0319, USA}
\author{K.\,L.~Campman, A.\,C. Gossard}
\affiliation{Materials Department, University of California at Santa Barbara, Santa Barbara, California 93106-5050, USA}

\begin{abstract}
\noindent We report on the observation of the Pancharatnam-Berry phase in a condensate of indirect excitons (IXs) in a GaAs coupled quantum well structure. The Pancharatnam-Berry phase leads to phase shifts of interference fringes in IX interference patterns. Correlations are found between the phase shifts, polarization pattern of IX emission, and onset of IX spontaneous coherence. The Pancharatnam-Berry phase is acquired due to coherent spin precession in IX condensate. The effect of the Pancharatnam-Berry phase on the IX phase pattern is described in terms of an associated momentum.
\end{abstract}

\pacs{73.63.Hs, 78.67.De, 05.30.Jp}

\date{\today}

\maketitle

The Pancharatnam-Berry phase is a phase appearing when the polarization state of light changes. It was discovered in studies of polarized light~\cite{Pancharatnam1956} and introduced as a topological phase for matter wave functions~\cite{Berry1984}. The connection between these two phases was established~\cite{Ramaseshan1986, Berry1987} and an analogy between the phase shift of polarized light and the Aharonov-Bohm effect was noted~\cite{Berry1987}. The Pancharatnam-Berry phase was measured for light in laser interferometers~\cite{Bhandari1988, Simon1988}. A variety of Pancharatnam-Berry phase optical elements were developed, including diffraction gratings~\cite{Bomzon2002, Hasman2003}, focusing lenses~\cite{Hasman2003}, and phase elements that create beams with an orbital angular momentum~\cite{Biener2002, Marrucci2006, Devlin2017}.

Excitons, bound pairs of electrons and holes in semiconductors, are matter waves that can directly transform to photons that inherit the excitons' energy, coherence, and polarization state. This makes excitons a unique interface between matter and light, and in turn,
a model system for exploring the Pancharatnam-Berry phase. Recent studies led to the discovery of polarization textures in light emission of indirect excitons (IXs)~\cite{High2012, High2013, Congjun2008, Matuszewski2012, Kyriienko2012, Vishnevsky2013, Kavokin2013, Sigurdsson2014, Bardyn2015, Li2016} and exciton-polaritons~\cite{Kavokin2005, Leyder2007, Cilibrizzi2015}. In this work, we report on experimental observation of the Pancharatnam-Berry phase in a condensate of IXs.

\begin{figure*}
\begin{center}
\includegraphics[width=17.5cm]{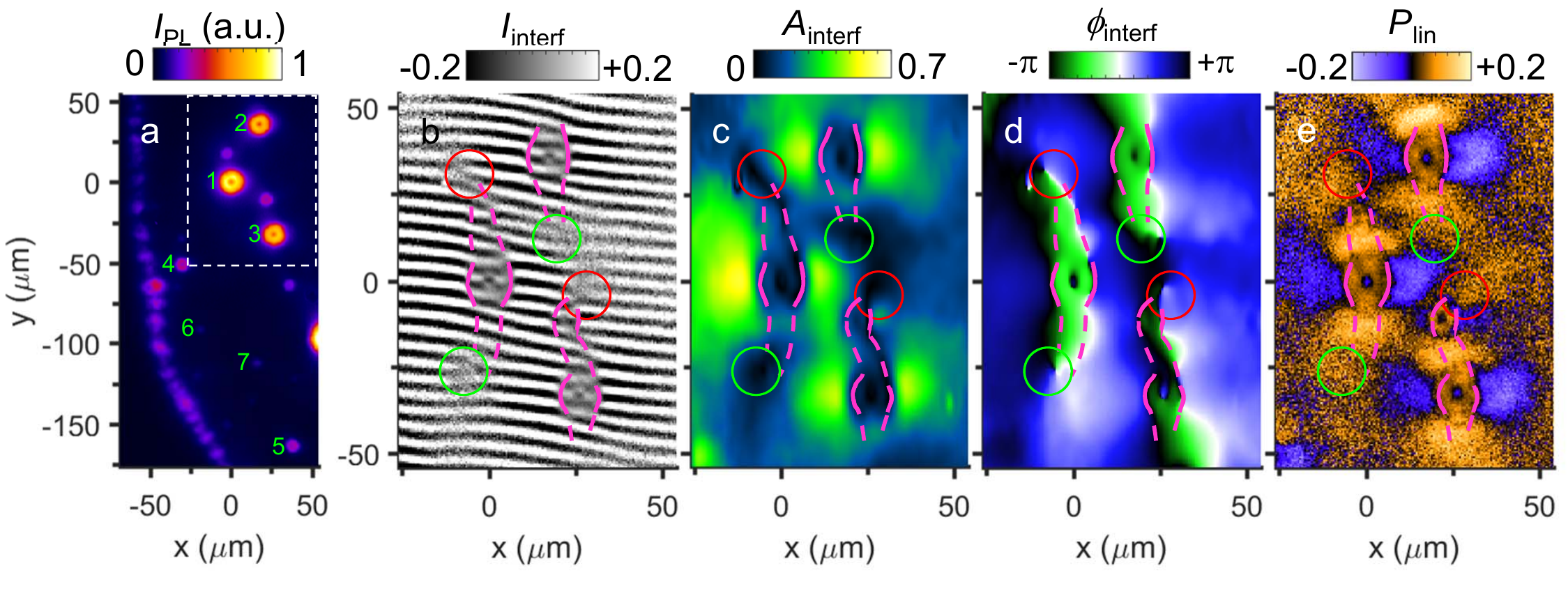}
\caption{\label{fig1} {\bf IX coherence and polarization patterns.} (a) IX emission image showing LBS 1 -- 7 numbered according to their emission power. The external ring is seen on the left. (b-d) Coherence and polarization patterns in the region of LBS 1 -- 3 marked by the dashed rectangle in (a). (b) Shift-interference pattern of IX emission, $I_{\rm interf}(x,y)$. The shift $\delta x = 2$~$\mu$m. (c,d) Amplitude (c) and phase (d) of interference fringes in (b), $A_{\rm interf}(x,y)$ and $\phi_{\rm interf}(x,y)$. (e) The linear polarization of IX emission, $P_{\rm linear}(x,y)$. In (b), the positions of phase shifts of interference fringes are marked by magenta lines and the positions of left (right) forks of interference fringes are marked by green (red) circles. The lines are solid in the circular region around each LBS where the phase shifts are sharp and dashed outside these regions where the phase shifts are smoother. These lines and circles are copied to (c-e) to show spatial correlations in $A_{\rm interf}(x,y)$, $\phi_{\rm interf}(x,y)$, and $P_{\rm linear}(x,y)$. Excitation power $P = 1.2$~mW. $T_{\rm bath} = 0.1$~K.}
\end{center}
\end{figure*}

An IX is a bound pair of an electron and a hole confined in spatially separated layers. IXs can be realized in coupled quantum well structures (CQW). The spatial separation reduces the overlap of the electron and hole wavefunctions, leading to IX lifetimes orders of magnitude longer than those of spatially direct excitons. IXs also form oriented dipoles and the IX dipole-dipole repulsion facilitates disorder screening. The long IX lifetimes and ability to screen disorder enable long-range transport of IXs before recombination~\cite{Hagn1995}, with transport distances reaching hundreds of microns. Furthermore, the reduction of electron-hole overlap leads to the suppression of the spin relaxation mechanism due to exchange interaction between the electron and hole and, in turn, to orders of magnitude enhancement of the IX spin relaxation time with respect to direct excitons in conventional quantum wells~\cite{Maialle1993}. This enables IX spin transport over substantial distances reaching $\sim 1$~$\mu$m for a classical IX gas~\cite{Leonard2009}.

Due to their long lifetimes IXs can cool well below the temperature of quantum degeneracy~\cite{Butov2001} and form a condensate in momentum ($k$) space~\cite{High2012}. IX condensation is detected by measurement of IX spontaneous coherence with a coherence length much larger than in a classical gas~\cite{High2012}. The large coherence length observed in an IX condensate, reaching $\sim 10$~$\mu$m, indicates coherent IX transport with suppressed scattering~\cite{High2012}, in agreement with theory~\cite{Lozovik1976}. The suppression of scattering results in the suppression of the Dyakonov-Perel and Elliott-Yafet mechanisms of spin relaxation~\cite{Dyakonov2008} and as a result, in a strong enhancement of the spin relaxation time in IX condensate. Long-range coherent exciton spin transport was observed in an IX condensate~\cite{High2013}. The coherent spin transport and spin precession create helical patterns in linear polarization and four-leaf patterns in circular polarization of IX emission~\cite{High2013}.

The polarization patterns~\cite{High2013} are observed in the region of the external ring and localized bright spot (LBS) rings in the IX emission pattern. These rings form on the boundaries of electron-rich and hole-rich regions created by current through the structure (specifically, by the current filament at the LBS center for an LBS) and optical excitation, respectively; see Ref.~\cite{Yang2015} and references therein. An LBS is a stable, well defined, and tunable source of cold IXs~\cite{Yang2015} thus an ideal system for studying coherence and polarization phenomena. In this work, we explore LBS to uncover the Pancharatnam-Berry phase in a condensate of IXs.

The experiments are performed on $n-i-n$ GaAs/AlGaAs CQW. The $i$ region consists of a single pair of 8-nm GaAs QWs separated by a 4-nm Al$_{0.33}$Ga$_{0.67}$As barrier and surrounded by 200-nm Al$_{0.33}$Ga$_{0.67}$As layers. The $n$ layers are Si-doped GaAs with Si concentration $5 \cdot 10^{17}$~cm$^{-3}$. The indirect regime where IXs form the ground state is realized by the voltage applied between $n$ layers. The small in-plane disorder in the CQW is indicated by the emission linewidth of 1~meV. IXs cool to temperatures within $\sim 50$~mK of the lattice temperature~\cite{Butov2001}, which was lowered to 100~mK in an optical dilution refrigerator. This cools IXs well below the temperature of quantum degeneracy, which is in the range of a few kelvin for typical IX density $10^{10}$~cm$^{-2}$~\cite{Butov2001}. The laser excitation is performed by a 633~nm HeNe laser. It is more than 400 meV above the energy of IXs and farther than 80~$\mu$m away from the studied region. Therefore, IX coherence and polarization are not induced by photoexcitation and form spontaneously. Because of their separation from the laser excitation spot where excitons are heated by the laser, LBS are sources of cold IXs.

Different LBS offer IX sources of different strength and spatial extension (Fig.~1a), furthermore these parameters can be controlled by optical excitation and voltage~\cite{Yang2015}. This variability gives the opportunity to measure correlations between coherence and polarization.

Figure~1b shows the interference pattern of IX emission $I_{\rm interf}(x,y)$ measured by shift-interferometry: The emission images produced by each of the two arms of the Mach-Zehnder interferometer are shifted with respect to each other to measure the interference between the emission of IXs separated by $\delta \bf r$ in CQW plane~\cite{Suppl}. Figures~1c and 1d show the amplitude $A_{\rm interf}(x,y)$ and phase $\phi_{\rm interf}(x,y)$ of interference fringes in Fig.~1b.

$A_{\rm interf}$ describes the degree of IX coherence. Spontaneous coherence of matter waves is equivalent to condensation of particles in momentum space. The Fourier transform of the first-order coherence function $g_1(\delta x)$ gives the particle distribution $n_k$ in the momentum space. In turn, the width of $g_1(\delta x)$, the coherence length $\xi$, is inversely proportional to the width of $n_k$. In a classical gas with $n_k$ given by the Maxwell-Boltzmann distribution, $\xi$ is close to the thermal de Broglie wavelength $\lambda_{\rm dB} = (2 \pi \hbar^2/mT)^{1/2}$ and is small ($\lambda_{\rm dB} \sim 0.5$~$\mu$m for IXs with $m = 0.22 m_0$ at $T = 0.1$~K).
The measurement of spontaneous coherence with $\xi \gg \lambda_{\rm dB}$ is a direct measurement of Bose-Einstein condensation.

The measured $A_{\rm interf}(\delta x)$ is given by the convolution of $g_1(\delta x)$ with the point-spread function (PSF) of the optical system~\cite{Fogler2008}. The PSF width corresponds to the spatial resolution ($\sim1.5$~$\mu$m in this experiment). For a classical IX gas, $g_1(\delta x)$ is narrow and $A_{\rm interf}(\delta x)$ fits well to the PSF, while for the IX condensate, $g_1(\delta x)$ and, in turn, $A_{\rm interf}(\delta x)$ extend to large $\delta x$~\cite{High2012}. While a more detailed picture is obtained by measuring $g_1(\delta x)$ as in~\cite{High2012}, mapping IX condensate can be done by measuring $A_{\rm interf}(x,y)$ at one value of $\delta x$ chosen to exceed both $\lambda_{\rm dB}$ and the PSF width. For such $\delta x$, a low $A_{\rm interf}$ is observed for a classical gas and a high $A_{\rm interf}$ -- for the condensate. For the parameters of our system, $\delta x = 2$~$\mu$m is optimal for this experiment.

The IX gas is classical close to the heating sources in the LBS central region (this heating is due to the current filament at the LBS center and the binding energy released at IX formation~\cite{Butov2004}). This is revealed by the small amplitude of the interference fringes $A_{\rm interf}$ at $r < r_{\rm coh}$ (Fig.~1c). IXs cool down away from the heating sources and approach the condensation temperature. At $r = r_{\rm coh}$, $A_{\rm interf}$ sharply rises, indicating the condensation.

Figures~1b and 1d show that the phase of interference fringes sharply changes at the distance from the LBS center $r = r_{\rm phase}$. The comparison of Figs.~1b, 1c, and 1d shows that the phase shifts occur at the same location as condensation, $r_{\rm phase} = r_{\rm coh}$.

\begin{figure}
\begin{center}
\includegraphics[width=8.5cm]{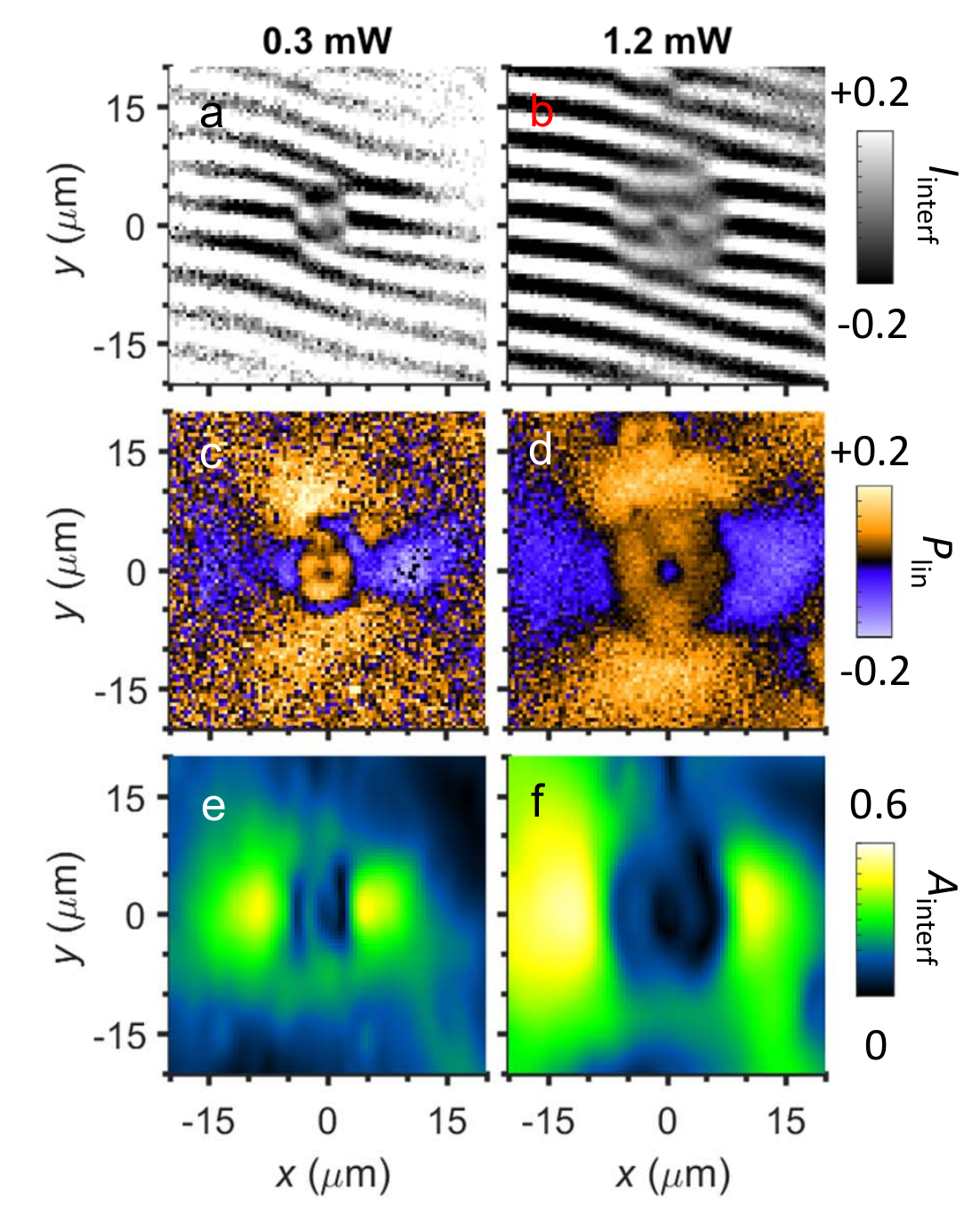}
\caption{\label{fig2}  {\bf IX coherence and polarization patterns for different excitation powers}. (a,b) Shift-interference pattern of IX emission. $\delta x = 2$~$\mu$m. (c,d) The linear polarization of IX emission. (e,f) Amplitude of interference fringes in (a,b). $P = 0.3$~(1.2)~mW for left (right) column. $T_{\rm bath} = 0.1$ K. The data correspond to LBS 1.}
\end{center}
\end{figure}

The phase shifts are sharp in the circular region around each LBS. However smoother phase shifts can be followed further (as shown by dashed magenta lines in Fig.~1b). The lines embracing the phase domains of interference fringes end by left- and right-forks of interference fringes at the opposite ends of the domain (the forks are shown by circles in Fig.~1b), indicating that the fork origin is related to the phase domains. This picture is observed for each LBS, although some patterns are complicated by the proximity of other LBS sources.

To uncover the origin of the phase shifts and associated phase domains of interference fringes, we compare their locations with the pattern of linear polarization of IX emission, $P_{\rm linear}(x,y)$ (Fig.~1e). The IX polarization patterns were observed in~\cite{High2012,High2013}. A ring of linear polarization is seen for each LBS in the region $r < r_{\rm linear}$ where the IX gas is classical, $r_{\rm linear} = r_{\rm coh}$ (compare Fig.~1c and 1e). The linear polarization originates from the distribution of IXs over the linearly polarized IX states~\cite{High2013}. A helical IX polarization texture winding by $2\pi$ around the origin, i.e. a vortex of linear polarization, emerges at $r > r_{\rm linear}$ where the IX condensate forms (Fig.~1e). IXs propagate away from the LBS source, therefore the IX linear polarization is perpendicular to IX momentum in this region. The polarization patterns are described by the theory based on coherent IX transport and coherent precession of spins of electrons and holes in IX condensate~\cite{High2012,High2013}.

The comparison of Figs.~1b-e shows that for all LBS sources, the phase shifts of interference fringes are observed when the polarization state of IX emission changes. To further examine this relationship, we measured IX coherence and polarization patterns at different laser exciation powers $P$. As in~\cite{Yang2015}, we also adjusted the applied voltage $V$ keeping the external ring radius constant. The simultaneous increase of $P$ and $V$ leads to the enhancement of both electron and hole sources and, as a result, the exciton source at each LBS.

Increasing the excitation power increases $r_{\rm phase}$ (Fig.~2a,b), $r_{\rm linear}$ (Fig.~2c,d), and $r_{\rm coh}$ (Fig.~2e,f). The increase of $r_{\rm coh}$ with $P$ follows an enhanced heating at the LBS central region due to the enhanced electron and hole sources. Figure~2 shows that $r_{\rm linear}$ stays equal to $r_{\rm coh}$ with increasing $P$, confirming that the polarization textures appear in the IX condensate, in agreement with theory~\cite{High2012,High2013}. Remarkably, Figure 2 shows that $r_{\rm phase}$ also keeps equal to $r_{\rm linear}$ with increasing $P$.

\begin{figure}
\begin{center}
\includegraphics[width=8.5cm]{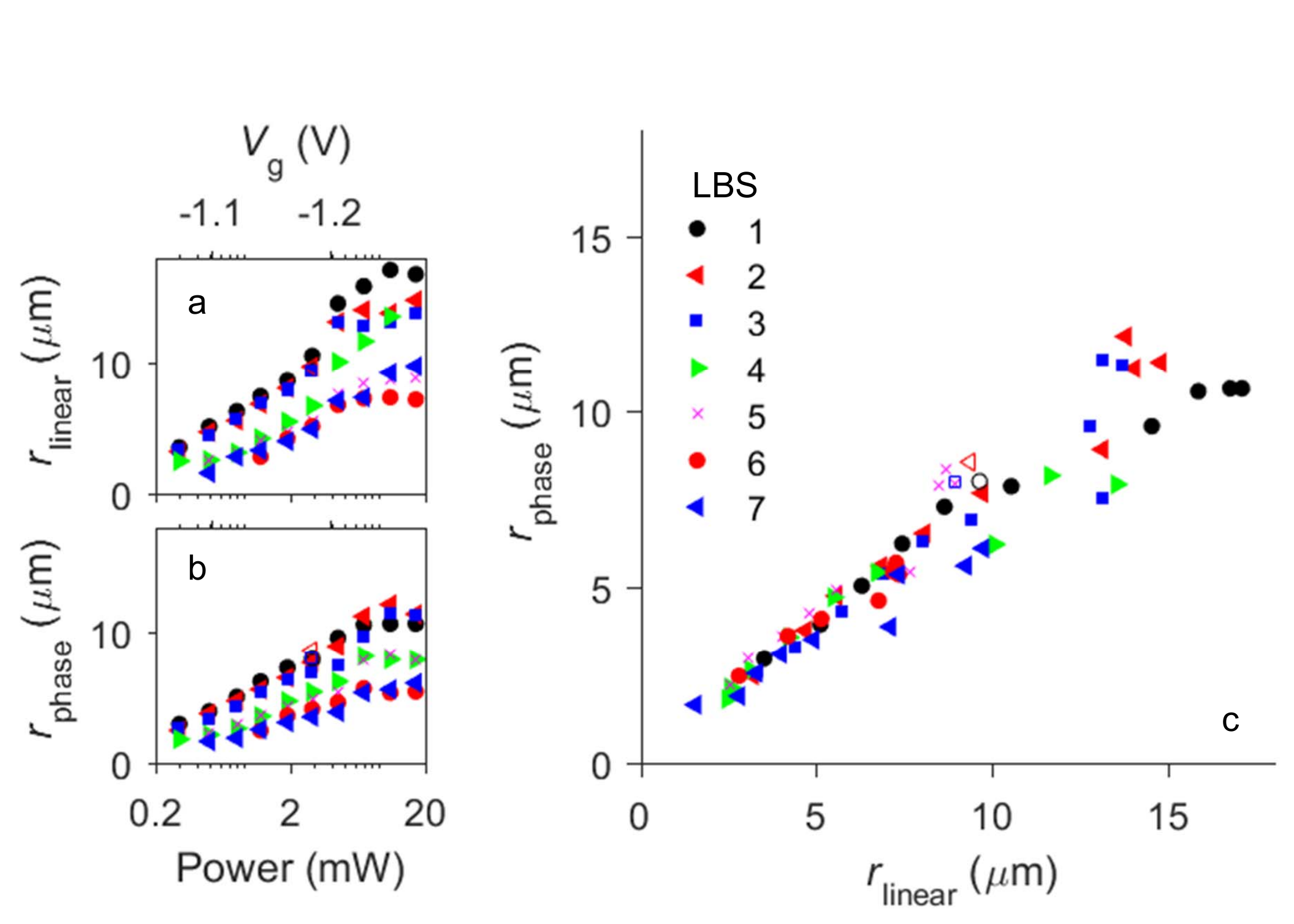}
\caption{\label{fig3} {\bf Correlation between the phase shifts and polarization pattern.} (a) $r_{\rm linear}$ for different LBS vs $P$. (b) $r_{\rm phase}$ for different LBS vs $P$. (c) $r_{\rm phase}$ vs $r_{\rm linear}$. Solid (open) symbols correspond to 2~$\mu$m shift in $\hat{x}$ ($\hat{y}$). $T_{\rm bath} = 0.1$ K. The data for different LBS and different $P$ collapse on a universal line $r_{\rm phase} \approx r_{\rm linear}$.}
\end{center}
\end{figure}

To show the universality of the relationship between $r_{\rm phase}$ and $r_{\rm linear}$ we measured them for many LBS over a broad range of excitation powers (Fig.~3). Large variations of both $r_{\rm phase}$ and $r_{\rm linear}$ are observed for different LBS and $P$ (Fig.~3a,b). However, $r_{\rm phase}$ always stays equal to $r_{\rm linear}$. The data for different LBS and different $P$ collapse on a universal line $r_{\rm phase} \approx r_{\rm linear}$ (Fig.~3c). Slight deviation of the slope of the line $r_{\rm phase}(r_{\rm linear})$ from 1 may be related to the calibration accuracy~\cite{Suppl}.

The phase of IX wave function $\psi({\bf r})$ acquired as IXs propagate from the origin can be derived from the interference pattern. The interference pattern in the shift-interferometry experiment with shift $\delta \bf r$ can be simulated using the formula $I_{\rm interf}({\bf r}) = \vert \psi({\bf r} - \delta {\bf r} / 2) + e^{iq_ty} \psi({\bf r}+\delta {\bf r} / 2) \vert^2$, where $q_t = 2\pi\alpha/\lambda$ sets the period of the interference fringes, $\alpha$ is a small tilt angle between the image planes of the interferometer arms, and $\lambda$ is the emission wavelength. Figures~4a-c show the simulated $I_{\rm interf}(x,y)$ for $\psi({\bf r}) = e^{i{\bf k} \cdot {\bf r}}$ and the shift in the $x$ direction $\delta x = 2$~$\mu$m for IXs radially propagating from the origin with a constant small $k$ (Fig.~4a), with a constant larger $k$ (Fig.~4b), and with the small $k$ at $r < r_{\rm k}$ and larger $k$ at $r > r_{\rm k}$ (Fig.~4c). For $k = 0$, the interference fringes are parallel lines separated by $D = \lambda / \alpha$. Deviations of fringes from their zero-$k$ positions are given by $\delta y_N = - D/(2\pi) \cdot k_x \delta x$ ($N$ is the fringe number). Therefore, the measurement of $\delta y_N (x,y)$ allows estimating ${\bf k}(x,y)$. Phase shifts of interference fringes, i.e. jumps in $\delta y_N$, correspond to jumps in $k$ (Fig.~4c). The values of $k$ at $r < r_{\rm k}$ and $r > r_{\rm k}$ in Fig.~4c were selected to qualitatively reproduce the measured $I_{\rm interf}$ in Fig.~1 and, in turn, illustrate a jump in $k$ at $r = r_{\rm phase}$.

\begin{figure}
\begin{center}
\includegraphics[width=8.5cm]{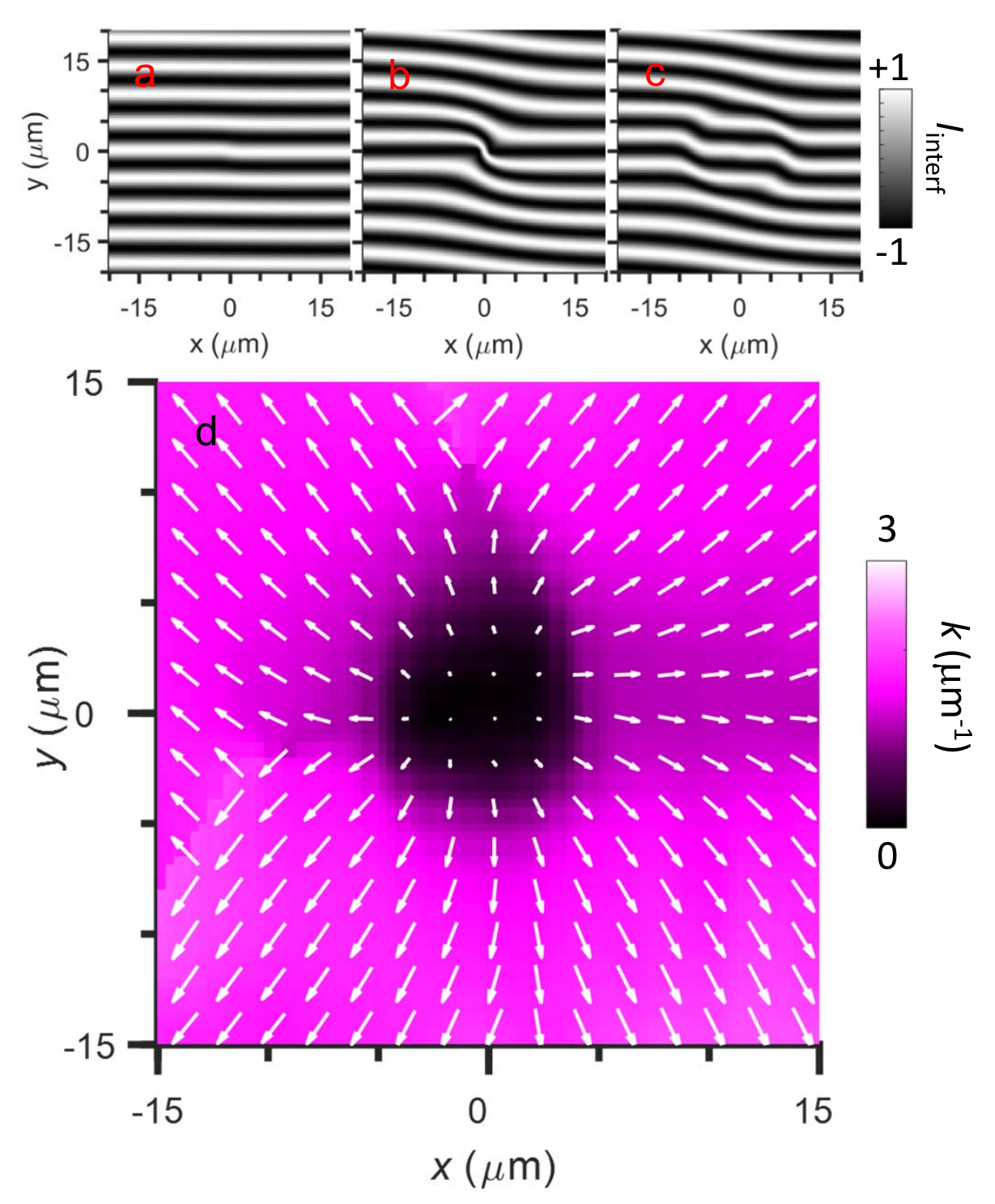}
\caption{\label{fig4} {\bf Spatial pattern of IX momentum $\bf k$.} (a-c) Simulation of IX shift-interference pattern for $k = 0.1$~${\mu}$m$^{-1}$ (a), for $k = 1.5$~${\mu}$m$^{-1}$ (b), and for $k = 0.1$~${\mu}$m$^{-1}$ at $r < r_{\rm k}$ and $k = 1.5$~${\mu}$m$^{-1}$ at $r > r_{\rm k}$ (c). $r_{\rm k} = 8$~$\mu$m. $\delta x = 2$~$\mu$m. IXs propagate away from the origin. (d) Pattern of $\bf k$ extracted from shift-interferometry measurements with 2~$\mu$m shift in $\hat{x}$ and $\hat{y}$. The direction and size of arrows indicate $\bf k$ direction and magnitude, respectively. $P = 2.9$~mW. $T_{\rm bath} = 0.1$~K. The data correspond to LBS 6.}
\end{center}
\end{figure}

The $x$ and $y$ components of $\bf k$ were derived within the approximation $\psi({\bf r}) = e^{i{\bf k} \cdot {\bf r}}$ by fitting the patterns of interference fringes measured for shifts in the $x$ and $y$ directions using $k_x (x,y) = - 2\pi/D \cdot \delta y_N(x,y)/\delta x$ and $k_y (x,y) = - 2\pi/D \cdot \delta y_N(x,y)/\delta y$, respectively. This allowed obtaining a map of IX momentum $\bf k$, describing the evolving IX phase (Fig.~4d). This map shows that IXs propagate away from the LBS source and the magnitude of $k$ is small at the origin close to the LBS center and sharply increases at $r = r_{\rm phase}$.

The experiment shows that the phase shifts correlate with the polarization pattern of IX emission and onset of IX spontaneous coherence. The correlation between the phase shift and polarization demonstrates the Pancharatnam-Berry phase of IXs. This phenomenon is discussed below.

For uncondensed IXs at $r < r_{\rm coh}$, the spin relaxation is fast and coherent spin precession is not observed. However, IX condensation at $r > r_{\rm coh}$ dramatically enhances the spin relaxation time leading to coherent spin precession and, in turn, precession of the polarization state of IX emission. This polarization precession generates the evolving Pancharatnam-Berry phase of IXs which is detected as the shift of the interference fringes.

Estimates for the Pancharatnam-Berry phase are presented below. When the polarization state of light goes along a closed contour on the Poincar\'{e} sphere the Pancharatnam-Berry phase acquired in this polarization cycle is equal to half the solid angle subtended by the contour at the center of the sphere, $\Omega/2$~\cite{Pancharatnam1956, Berry1984, Ramaseshan1986, Berry1987, Bhandari1988, Simon1988, Bomzon2002, Hasman2003, Biener2002, Marrucci2006, Devlin2017}.

To demonstrate the concept, we simulated the polarization evolution within the model of IX spin precession developed in~\cite{High2012, High2013}. This model describes the polarization patterns of IX emission~\cite{High2013}. Here, we use the same electron and hole spin-orbit interaction constants and splittings between four IX states (with spin projections $J_z = \pm 2, \pm 1$) as in~\cite{High2013}. These parameters were obtained to fit the IX polarization patterns in~\cite{High2013}. At the same time, they demonstrate a concept of the Pancharatnam-Berry phase acquired due to the IX spin precession. The initial polarization in the simulations is taken as horizontal to qualitatively follow the experiment (Fig.~1e).

\begin{figure}
\begin{center}
\includegraphics[width=8.5cm]{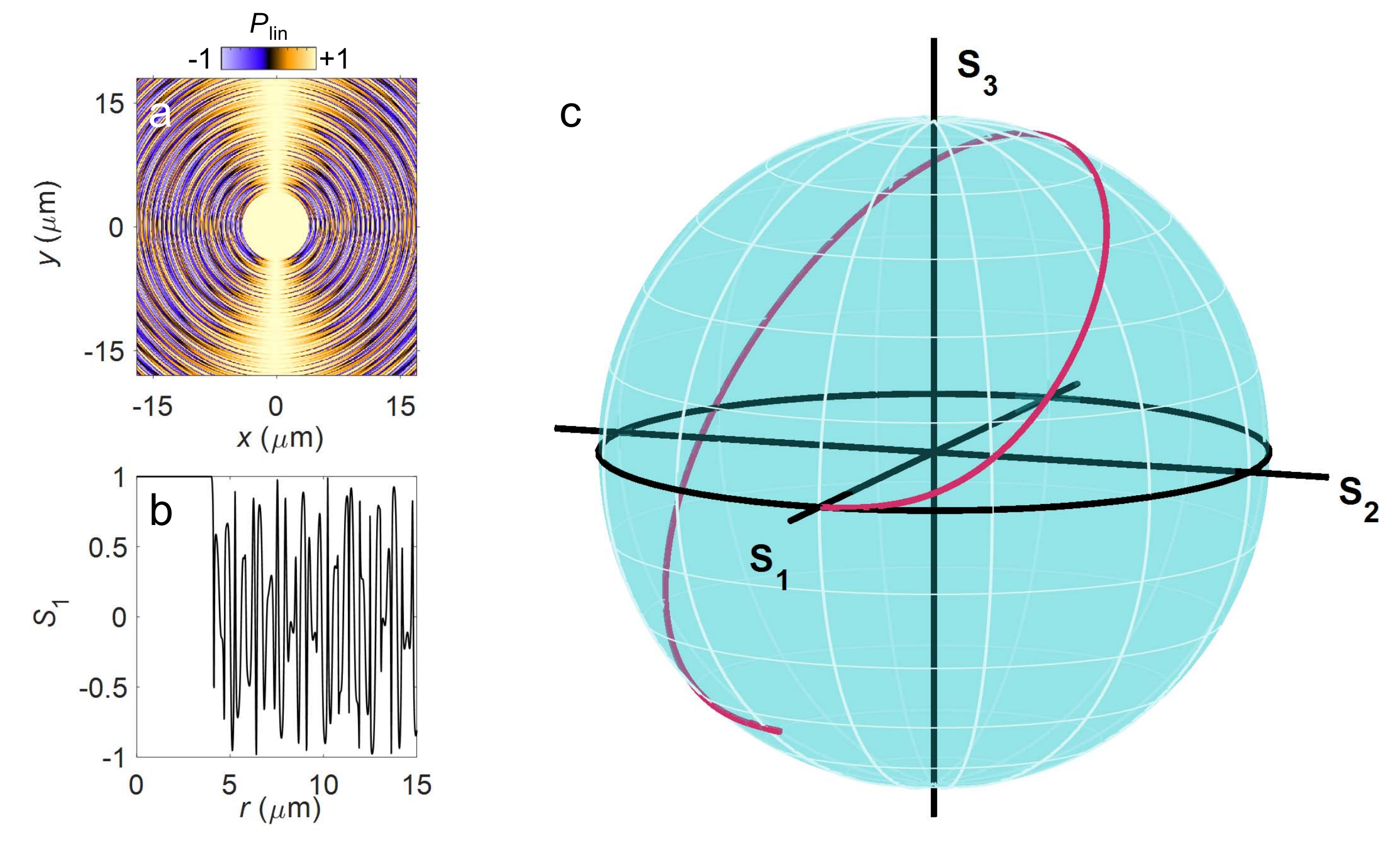}
\caption{\label{fig5} {\bf Simulation of IX polarization state.} (a) Simulated S1 Stokes' vector component corresponding to linear polarization of IX emission. Initial conditions at $r = r_{\rm p}$ correspond to linear polarization. Oscillatory S1 behaviour due to coherent spin precession occurs for $r > r_{\rm p}$.  $r_{\rm p} = 4$~$\mu$m. (b) Diagonal cross-section of (a). (c) IX polarization state on Poincar\'{e} sphere for one fast polarization oscillation cycle in (a).}
\end{center}
\end{figure}

The simulated S1 component of Stokes' vector corresponding to linear polarization of IX emission is presented in Fig.~5a,b. The polarization shows an oscillatory behaviour. Its slow long-length component is responsible for the polarization pattern shown in Fig.~1e and studied earlier in~\cite{High2012, High2013}. The fast short-length component has the spatial period $\sim 0.3$~$\mu$m and is not resolved with 1.5~$\mu$m optical resolution in the imaging experiment. However, these fast changes of the polarization state generate the evolving Pancharatnam-Berry phase of IXs. Figure~5c shows the simulated IX polarization state on the Poincar\'{e} sphere for one fast polarization oscillation cycle in Fig.~5a. The IX polarization state goes over a nearly closed contour on the Poincar\'{e} sphere. The Pancharatnam-Berry phase acquired by IXs over this nearly closed contour can be roughly estimated by connecting the initial and final points and calculating $\Omega/2$ of the obtained contour (a correction given by half the solid angle of the triangle build on the initial and final polarization states and the analyser polarization state is small for nearly closed contours)~\cite{Pancharatnam1956}. In turn, a momentum $k_{\rm PB}$ associated with the acquired Pancharatnam-Berry phase can be estimated as $\sim \Omega / (2l)$ where $l$ is the IX path passed during the polarization cycle. For $\Omega/2 \sim \pi/2$~(Fig.~5c) and $l \sim 0.3$~$\mu$m (Fig.~5a,b), this estimate gives $k_{\rm PB} \sim 5$~$\mu$m$^{-1}$. This rough estimate is of the same order of magnitude as the jump in IX momentum (Fig.~4d), which occurs when the coherent spin precession generating the evolving Pancharatnam-Berry phase starts in IX condensate.

We note that the polarization contour and $l$ depend on a number of parameters including the electron and hole spin-orbit interaction constants, splittings between IX states, their occupation, IX momentum, and initial polarization. Furthermore, the employed model~\cite{High2012, High2013} can be refined to include for instance interaction, damping, and recombination. We demonstrate here a concept of the Pancharatnam-Berry phase and its effect on interference patterns. The parameter fitting and model refining form the subject for future works.

In summary, shift-interferometry and polarization imaging of IX emission show that the phase shifts of interference fringes correlate with polarization pattern of IX emission and onset of IX spontaneous coherence. The correlation demonstrates the Pancharatnam-Berry phase in a condensate of IXs.

\section*{Acknowledgments}
We thank Lu Sham and Congjun Wu for discussions.
These studies were supported by DOE Office of Basic
Energy Sciences under award DE-FG02-07ER46449 and
NSF Grant No. 1640173 and NERC, a subsidiary of SRC,
through the SRC-NRI Center for Excitonic Devices.

\clearpage

\section*{Supplementary materials}

\noindent Figure~S1 shows the energy diagram of the coupled quantum well structure. Figure~S2 shows the interferometer setup. Figure~S3 is similar to Fig.~1 in the main text except Fig.~S3 also shows data without markup.

\renewcommand{\thefigure}{S\arabic{figure}}
\setcounter{figure}{0}

\begin{figure}[!htb]
\begin{center}
\includegraphics[width=4cm]{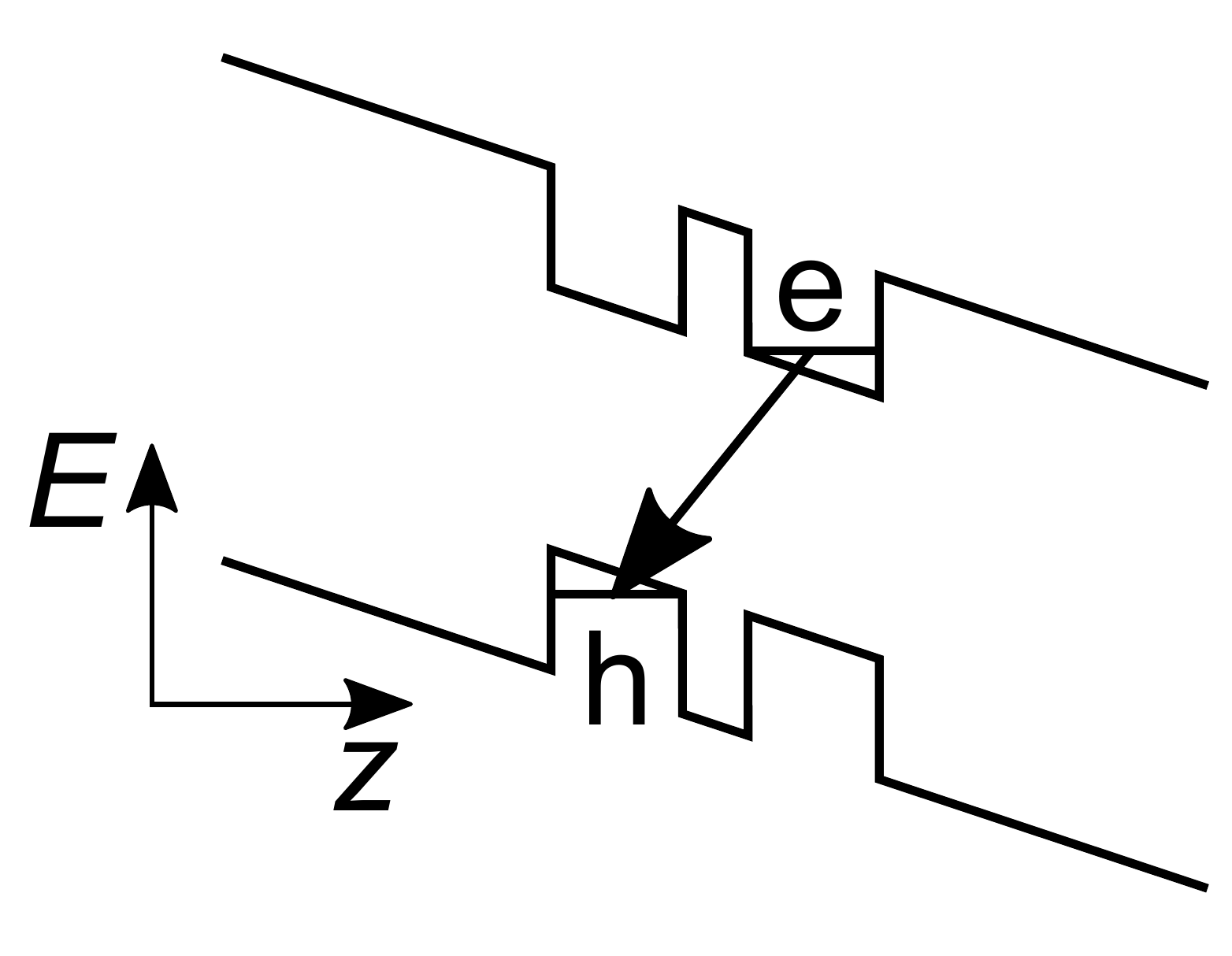}
\caption{\label{figS1} {\bf IX band diagram.} Energy diagram showing coupled quantum well structure with an applied electric field along the $z$-axis. e (h) indicate an electron (hole). The arrow indicates an indirect exciton.}
\end{center}
\end{figure}

\begin{figure}[!htb]
\begin{center}
\includegraphics[width=8.5cm]{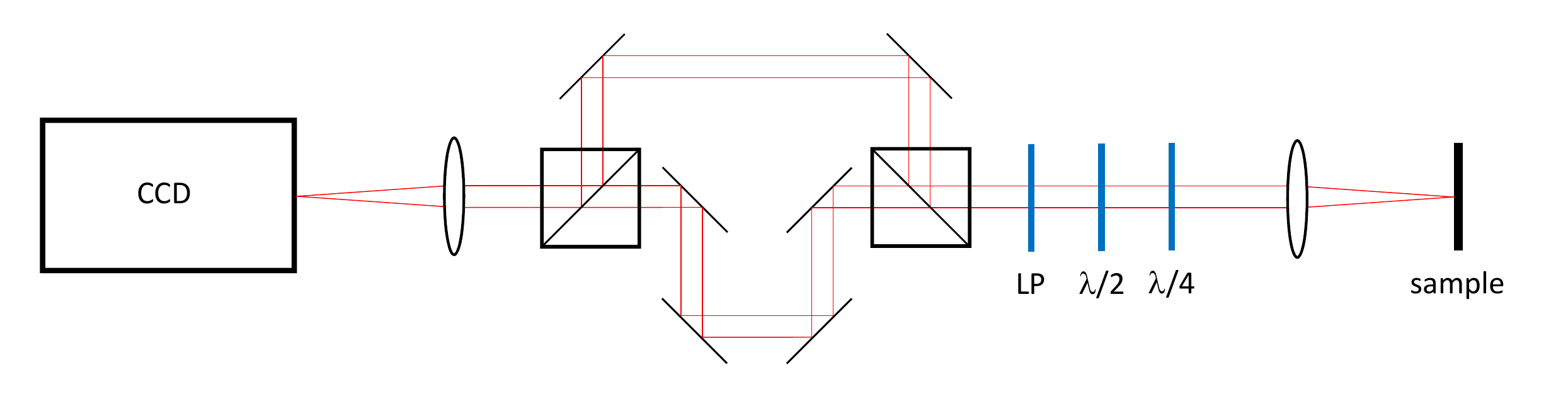}
\caption{\label{figS2} {\bf Imaging Setup.} Schematic of the
shift-interferometry and polarization imaging setup.}
\end{center}
\end{figure}

\begin{figure*}[!htb]
\begin{center}
\includegraphics[width=17cm]{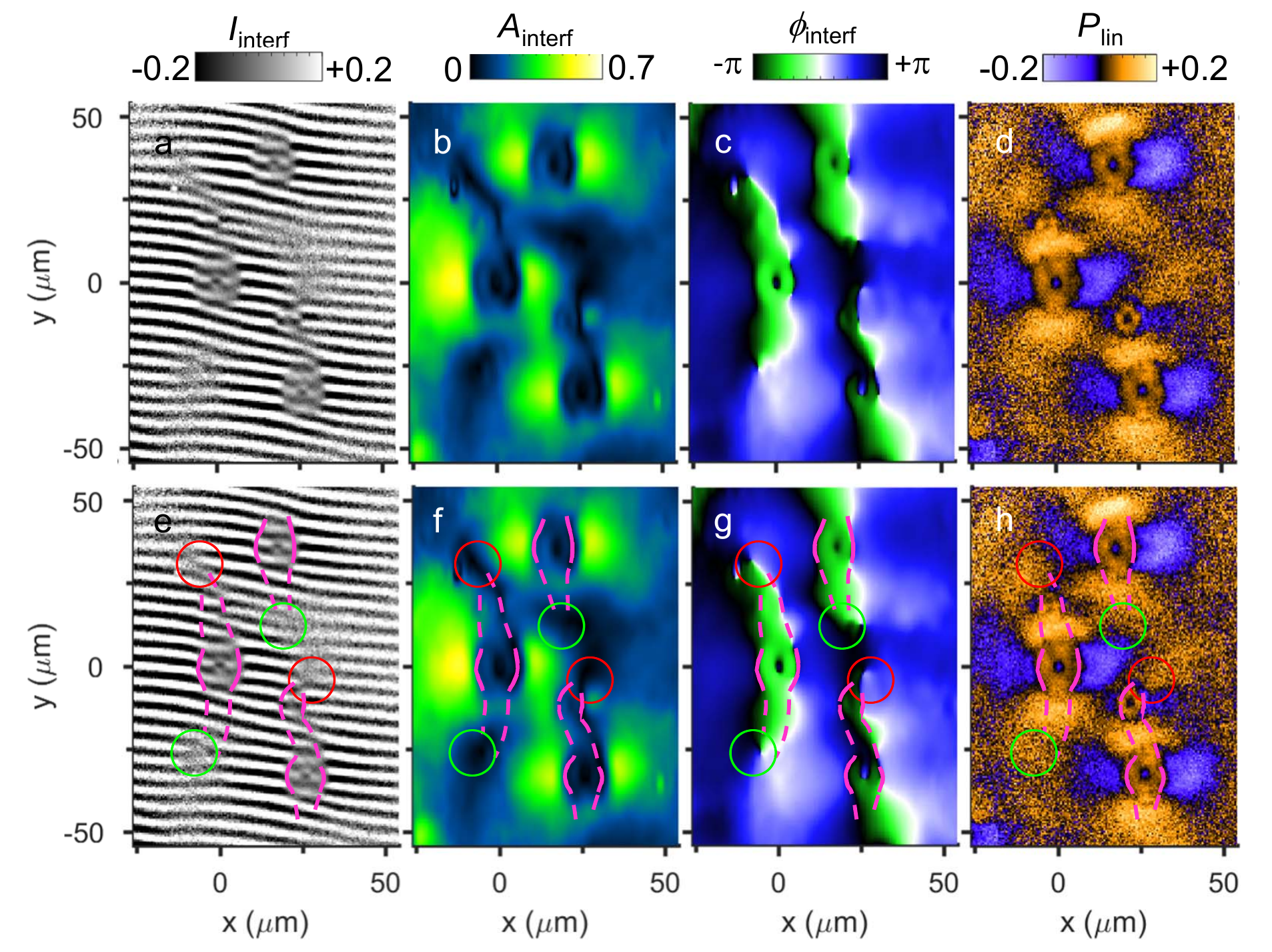}
\caption{\label{figS3} {\bf Coherence and polarization patterns of indirect excitons (IXs).} (a,e) Shift-interference pattern of IX emission, $I_{\rm interf}(x,y)$. The shift $\delta x = 2$~$\mu$m. (b,c,f,g) Amplitude, $A_{\rm interf}(x,y)$, (b,f) and phase, $\phi_{\rm interf}(x,y)$, (c,g) of interference fringes in (a,e). (d,h) The linear polarization of IX emission, $P_{\rm linear}(x,y)$. (a-d) show the raw data without markup. In (e), the positions of phase jumps of interference fringes are marked by magenta lines and the positions of left (right) forks of interference fringes are marked by green (red) circles. The lines are solid in the circular region around each LBS where the jumps are sharp and dashed outside these regions where the jumps are smoother. These lines and circles are copied to (f-h) to show spatial correlations in $A_{\rm interf}(x,y)$, $\phi_{\rm interf}(x,y)$, $P_{\rm linear}(x,y)$. Excitation power $P = 1.2$~mW. $T_{\rm bath} = 0.1$~K.}
\end{center}
\end{figure*}

\paragraph*{Experimental Setup}
We use a Mach-Zehnder (MZ) interferometer to probe the coherence of the exciton system (Fig.~S2). The emission beam is made parallel by an objective inside the optical dilution refrigerator and lenses. A combination of a quarter-wave plate and a half-wave plate converts the measured polarization of the emission to the $y$-polarization, which is then selected by a linear polarizer. This ensures only $y$-polarized light enters the MZ interferometer eliminating polarization-dependent effects in the interferometer and spectrometer.
The emission is split between arms of the MZ interferometer. The path lengths of the arms are equal. The interfering emission images produced by arm 1 and 2 of the MZ interferometer are shifted relative to each other along $x$ (or $y$) directions to measure the interference between the emission of excitons, which are laterally separated by $\delta x$ (or $\delta y$).
After the interferometer, the emission is filtered by an interference filter of linewidth $\pm 5$~nm adjusted to the exciton emission wavelength $\approx$~800nm. The filtered signal is focused to produce an image, which is measured by a liquid-nitrogen cooled CCD. We measure exciton emission intensity $I_{\rm 1}$ for arm 1 open, intensity $I_{\rm 2}$ for arm 2 open, and intensity $I_{\rm 12}$ for both arms open, and then calculate
\begin{equation}\label{eq:Iinterf}
I_{\rm interf} = \frac{I_{\rm 12} - I_{\rm 1} - I_{\rm 2}}{2\sqrt{I_{\rm 1}I_{\rm 2}}}
\end{equation}
shown in Fig.~S3a.
In general, for two partially coherent sources located at $\bf r_1$ and $\bf r_2$, one has the relation \cite{Milonni},
\begin{equation}\label{eq:coh}
I_{\rm interf} = \cos {\delta}\theta \left({\bf r}_{\rm 1}, {\bf r}_{\rm 2} \right)
\zeta \left({\bf r}_{\rm 1}, {\bf r}_{\rm 2} \right),
\end{equation}
where ${\delta}\theta \left({\bf r}_{\rm 1}, {\bf r}_{\rm 2} \right)$ is the phase difference of the two sources and $\zeta \left({\bf r}_{\rm 1}, {\bf r}_{\rm 2} \right)$ is their degree of coherence. In our experimental geometry, there is a small tilt angle $\alpha$ between the image planes of the two arms. As a result, the phase difference
\begin{equation}\label{eq:theta}
{\delta}\theta \left({\bf r}_{\rm 1}, {\bf r}_{\rm 2} \right) = q_{\rm y}y + \phi\left({\bf r}_{\rm 1}, {\bf r}_{\rm 2} \right)
\end{equation}
has a component linear in $y$ - the coordinate in the direction perpendicular to the tilt axis - which produces periodic oscillation of $I_{\rm interf}$. The period of the interference fringes is set by $q_{\rm y} = 2\pi\alpha/\lambda$. The coherence function $\zeta \left({\bf r}_{\rm 1}, {\bf r}_{\rm 2} \right)$ for ${\bf r}_{\rm 1}-{\bf r}_{\rm 2}= {\delta}{\bf r}$ is given by the amplitude of these interference fringes.
The amplitude and phase of the interference fringes in Fig.~S3 (b) and (c) were calculated from the interference pattern (Fig.~3(a)) using a two dimensional windowed fourier transform with a gaussian window.

At high temperatures, the linear polarization $P_{\rm linear} = \left(I_{\rm x}-I_{\rm y}\right)/\left(I_{\rm x}+I_{\rm y}\right)$ of exciton emission is expected to vanish, and indeed for $T_{\rm bath} = 7$K, in the region of an LBS, $P_{\rm linear}$ is small, $\lesssim 5\%$,
that is within the polarization calibration accuracy. $r_{\rm linear}$ is defined as the radius where $P_{\rm linear}$ along $\hat x$ changes sign.
The slope of the line $r_{\rm phase}(r_{\rm linear})$ (Fig.~3c in the main text) is equal to 1 within the calibration accuracy.

\end{document}